# Trends, Challenges, and Future Directions in Deep Learning for Glaucoma: A Systematic Review

**Authors:** Mahtab Faraji[123], Homa Rashidisabet[123], George R. Nahass [123], RV Paul Chan[13], Thasarat S Vajaranant[13], Darvin Yi[13*]

[1] *Illinois Eye and Ear Infirmary, Department of Ophthalmology and Visual Sciences, University of Illinois Chicago, Chicago, Illinois, United States*

[2] *Richard and Loan Hill Department of Biomedical Engineering, University of Illinois at Chicago*

[3] *Artificial Intelligence in Ophthalmology (Ai-O) Center, University of Illinois Chicago*

**Corresponding author:** Darvin Yi[13] , Email: dyi9@uic.edu

**Financial Support:** This work is supported by an Unrestricted Grant from Research to Prevent Blindness, Cless Family Foundation, NIH P30 EY001792 Core Grant.

## Abstract

Here, we examine the latest advances in glaucoma detection through Deep Learning (DL) algorithms using Preferred Reporting Items for Systematic Reviews and Meta-Analyses (PRISMA). This study focuses on three aspects of DL-based glaucoma detection frameworks: input data modalities, processing strategies, and model architectures and applications. Moreover, we analyze trends in employing each aspect since the onset of DL in this field. Finally, we address current challenges and suggest future research directions.

## 1. Introduction

Glaucoma is the second most common cause of blindness worldwide and a significant contributor to vision impairment globally, expected to affect 112 million patients by 2040 [1,2]. Glaucoma is a complex disease due to its lack of symptoms in the early disease stages, which leads to late detection and irreversible vision loss [3]. The various risk factors and complex genetics underlying glaucoma further complicate early diagnosis and effective treatment, thus underscoring the necessity of advanced and effective screening methods [3]. Since DL is becoming increasingly popular in the medical imaging field [4,5], several studies [6–10] have attempted to address these challenges by developing computer-aided diagnostic systems based on DL algorithms. These systems utilize DL methods for early and automated disease detection.

In light of this, several review papers [11–14] have surveyed these studies to summarize recent developments and explore remaining challenges in the field. However, to the best of our knowledge, no study has evaluated DL-based glaucoma detection literature concurrently on the following criteria:

1. Input data modalities: Color Fundus Photography (CFP), Optical Coherence Tomography (OCT), Visual Field (VF), and multi-modality.
2. Processing strategies: single- and two-step
3. Model architectures and applications: Convolutional Neural Networks (CNNs), Generative Adversarial Networks (GANs), and attention models, along with their applications in glaucoma detection, including classification, progression prediction, and image synthesis.
4. Trends in the utilization of 1, 2, and 3 since the DL onset in the field.

The structure of this paper is organized as follows: Section 2 details the methodology of our literature search methodology, and section 3 explores DL-based glaucoma detection frameworks as described by the taxonomy in Fig 1. The challenges encountered and potential future research directions are discussed in section 4, and section 5 provides our main conclusions and final thoughts.

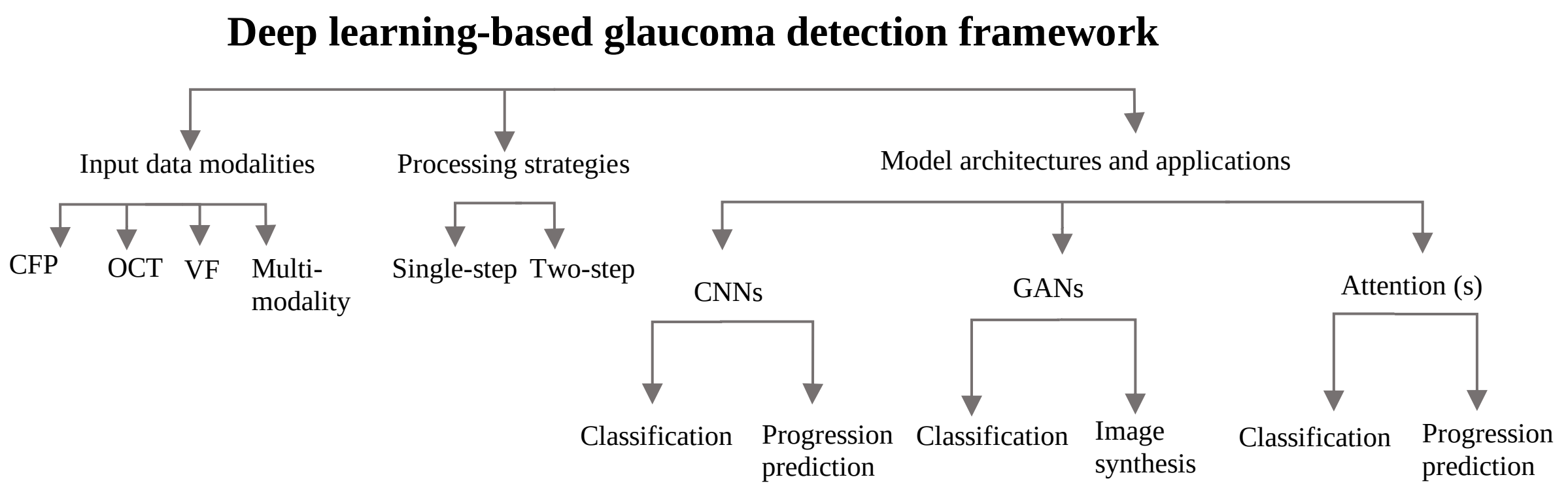

Fig 1 The taxonomy of the DL framework explored in this survey.

## 2. Literature search methodology

We conducted a systematic literature review following PRISMA guidelines (Fig 2), assessing different input data modalities, processing strategies, and model architectures and applications in glaucoma detection from 2016 to 2023.

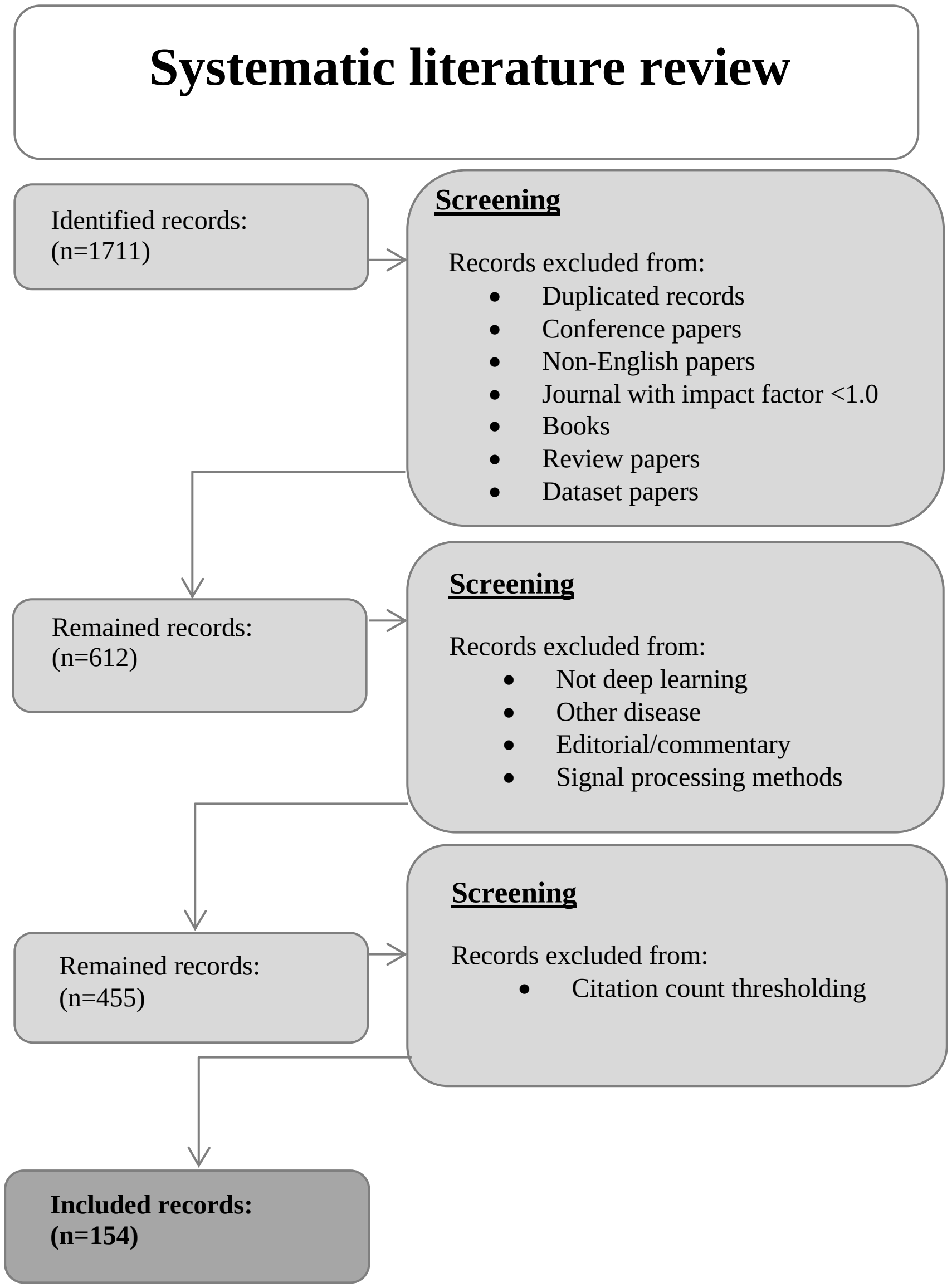

Fig 2: The overview of the PRISMA search methodology.

A comprehensive search of PubMed, Science Direct, and Scopus was performed using various search queries, including terms related to glaucoma and DL up to 2023. Our selection approach initially produced 1,711 records, then narrowed down to 455 records directly pertinent to DL in glaucoma detection. The remaining records were further filtered to include 154 final records by prioritizing studies with significant academic impact based on citation count thresholds, as shown in Table 1.

Table 1. Thresholding criteria for including/excluding literature based on the mean citation counts.

| | Year | Total ($N_y$) | Threshold ($T_y$) | Excluded literature | Included literature |
|---|---|---|---|---|---|
| | 2016 | **1** | - | - | 1 |
| | 2017 | **3** | - | - | 3 |
| | 2018 | **23** | 166.68 | 17 | 6 |
| | 2019 | **51** | 73.15 | 39 | 12 |
| | 2020 | **79** | 32.39 | 60 | 19 |
| | 2021 | **119** | 15.20 | 92 | 27 |
| | 2022 | **123** | 4.5 | 93 | 30 |
| | 2023 | **56** | - | - | 56 |
| **Total** | - | **455** | - | **301** | **154** |

The thresholds were determined by calculating the mean number of citations received by papers each year from 2018 to 2022 (See Eq.1). $C_{y,i}$ is the number of citations for the paper "i" published in a year "y", $N_y$ is the total number of papers published in a year "y," and, $T_y$ is the mean citation threshold for a year "y." It is significant to note that we excluded the years 2016 and 2017 from our analysis due to the emerging stage of research in those initial years and 2023 because of the recent status of publications. Studies with a citation count below the threshold value were excluded.

$$T_y = \frac{1}{N_y} \sum_{i=1}^{N_y} C_{y,i} \qquad for\ y = 2018, 2019, 2020, 2021, 2022 \qquad \text{Eq (1)}$$

The citation frequency by year for DL-based glaucoma detection studies reviewed in this survey is provided in Appendix A (Fig A.1). While we aimed to cover a comprehensive and representative collection of research, we acknowledge the potential accidental omission of some studies. As the concluding phase, we extracted relevant information from the 154 included records, encompassing details on input data modalities, processing strategies, model architectures, applications, and performance metrics.

# 3. Deep learning-based glaucoma detection framework

## 3.1. Input data modalities

The primary input modalities used in the DL-based glaucoma detection studies were CFP, OCT, VF, and multimodal CFP and OCT images, which offer structural information about the retina, whereas VF data provide functional insights [15]. This section outlines these modalities, dataset availability, and usage trends.

**CFP**: A CFP image (Fig 3(a)) is an essential part of ophthalmic examinations, providing detailed images of the size, shape, and color of critical regions of the retina. It includes retinal structures such as the optic disc (OD), optic cup (OC), blood vessels, and the neuroretinal rim [16]. Glaucoma can enlarge OC due to nerve fiber loss. This results in an increased Cup-to-Disc Ratio (CDR), a hallmark diagnostic feature of glaucoma, illustrated in Fig 3(b).

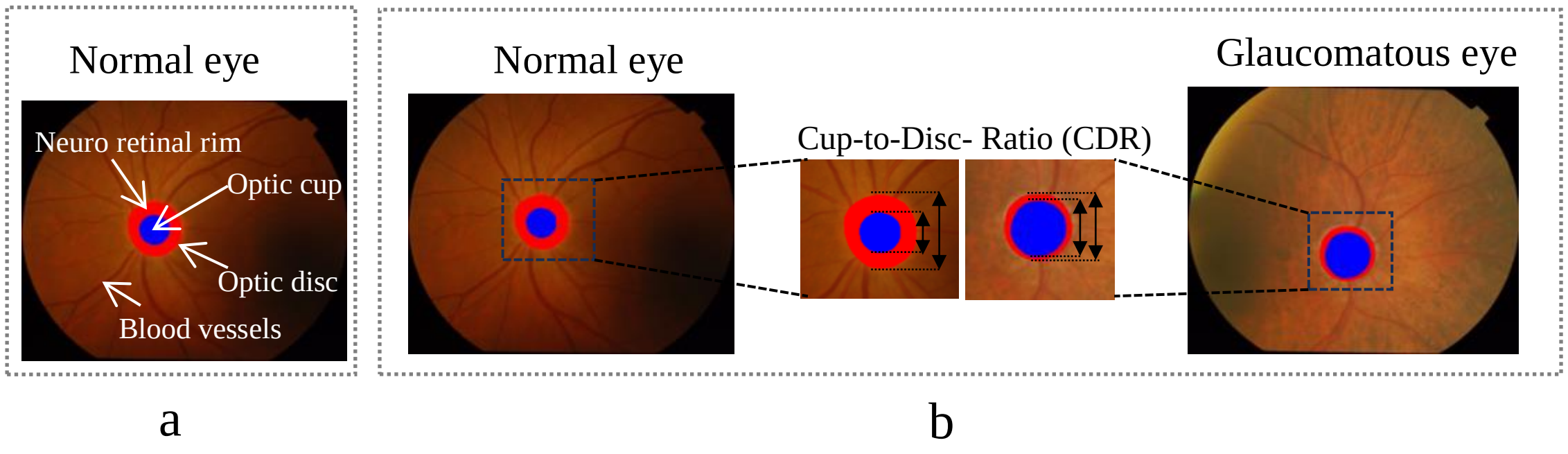

Fig 3. The comparative analysis of retinal structures in CFP image data. (a) A normal eye and detailed anatomical retinal structures; (b) A side-by-side representation of normal and glaucomatous eyes with CDR highlighted. The data is taken from the DRISHTI-GS publicly available dataset [17,18]

Table 2 shows a growing trend in the utilization of CFP images for DL-based glaucoma detection in the literature, with the majority of studies relying on publicly available datasets, such as RIM-ONE [19], MESSIDOR [20], REFUGE [21], RIGA [22], HRF [23], and ODIR [24]. A summary of some publicly available datasets for glaucoma detection is provided in Appendix B (Table B.1).

Table 2. Annual trend and data availability of different input data modalities in glaucoma detection studies.

| | **Annual trend** | | | | | | | | **Data availability** | | | |
|---|---|---|---|---|---|---|---|---|---|---|---|---|
| **Modality** | 2016 | 2017 | 2018 | 2019 | 2020 | 2021 | 2022 | 2023 | Public | Private | Public and private | **Total** |
| CFP | - | 2 | 6 | 9 | 11 | 18 | 18 | 35 | 55 | 27 | 17 | 99 |
| OCT | - | 1 | - | 2 | 5 | 5 | 7 | 11 | - | 30 | 1 | 31 |
| Multi-modality | - | - | - | 1 | 3 | 3 | 2 | 7 | - | 16 | - | 16 |
| VF | 1 | - | - | - | - | - | 4 | 2 | - | 9 | 1 | 10 |

**OCT:** OCT imaging has significantly advanced glaucoma diagnosis through its ability to capture detailed, cross-sectional images of the retina that are not visible when using modalities such as CFPs [25,26]. In particular, OCT provides comprehensive 3D images and quantitative measurements of various retinal layers, such as the Retinal Nerve Fiber Layer (RNFL), Ganglion Cell Complex (GCC), Retinal Pigmented Epithelium (RPE), and Inner Limiting Membrane (ILM) [27,28]. ILM and RPE are used to calculate CDR for glaucoma diagnosis. The segmentation of these two layers in glaucoma and normal OCT B-scan cases can be seen in Fig 4.

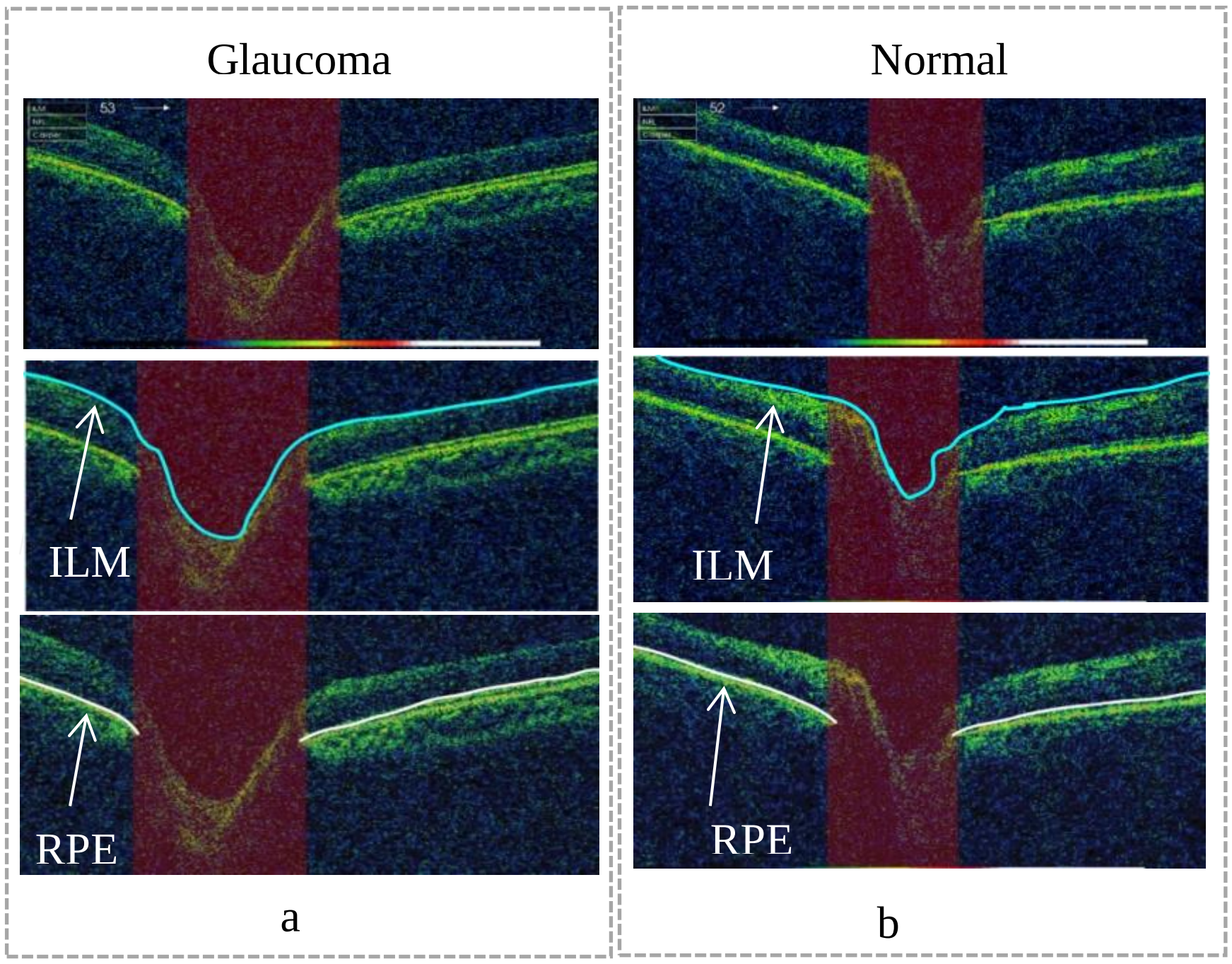

Fig 4. The comparative analysis of retinal structures in OCT image data [29]. (a) glaucomatous eye; (b) normal eye. The scans are taken from the publicly available Data on OCT and CFP Images [1].

Various OCT sub-modalities, such as OCT 2D B-scans [30,31], OCT 3D volumetric scans [32], AS-OCT [33,34], SS-OCT [35,36], and OCTA [37,38], have been used in glaucoma detection studies reviewed in this survey. These studies indicate a growing trend, with most studies relying on private datasets (Table 2).

[1] https://data.mendeley.com/datasets/2rnnz5nz74/2

**VF data:** The VF data is vital for identifying glaucoma-related defects and assessing retinal sensitivity loss [39–41]. In the VF test, shown in Fig 5(a), a patient responds to visual stimuli at various locations. Specific VF deficits are consistent with loss of vision due to glaucoma, providing valuable information to clinicians for accurate diagnosis (Fig 5(b)) [42].

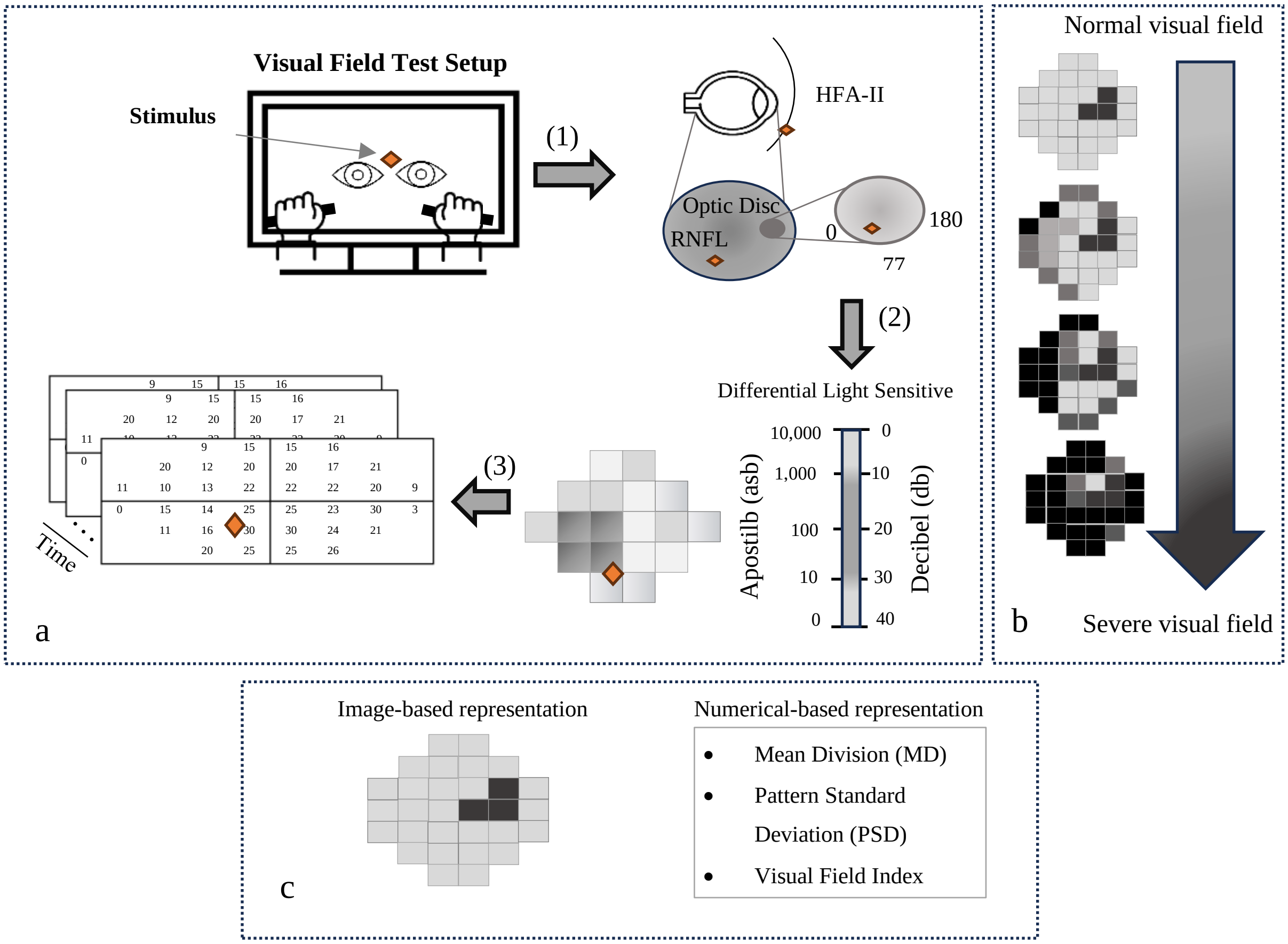

Fig 5. Visual field evaluation process [43,44]. (a) visual field-testing ;(b) VF patterns from normal eye to severe visual loss; (c) two types of VF inputs used in the DL-based glaucoma detection studies.

In glaucoma detection studies, the data from VF tests are presented in two main formats: image (e.g., gray-scale image) [45,46] and numerical representations (e.g., Mean Deviation (MD), Visual

Field Index (VFI) and Pattern Standard Deviation (PSD)) [47] (Fig 5(c)). Studies performing glaucoma detection using VF indicate a subtle but growing trend, with most studies relying on private datasets (Table 2).

**Multi-Modality data:** Different types of ophthalmic data, such as CFP, OCT, and VF, provide unique information on retinal pathology. The combination of any or all of these modalities, termed multi-modality, allows the unique features intrinsic to each data modality to be leveraged, improving the understanding of structural and functional changes in glaucoma [48]. Multi-modality is also defined as the combination of ophthalmic images and clinical data, such as demographics and medical history.

MMultimodalapproaches to glaucoma detection are becoming increasingly popular (Table 2) [42,49–51] due to a growing awareness of the value of combining several diagnostic methods [52]. Furthermore, it is noted that the datasets used in these studies are mainly private.

Robust data fusion techniques are essential to advance multimodal DL approaches for glaucoma detection. Data fusion techniques, such as early and late fusion (Fig 6), incorporate information from various modalities. The early fusion process, also known as feature level fusion, involves combining multiple input modalities into a feature vector prior to being fed into a machine learning or DL model for training [53]. Conversely, late fusion refers to the process of using predictions from multiple models to reach a final decision[53].

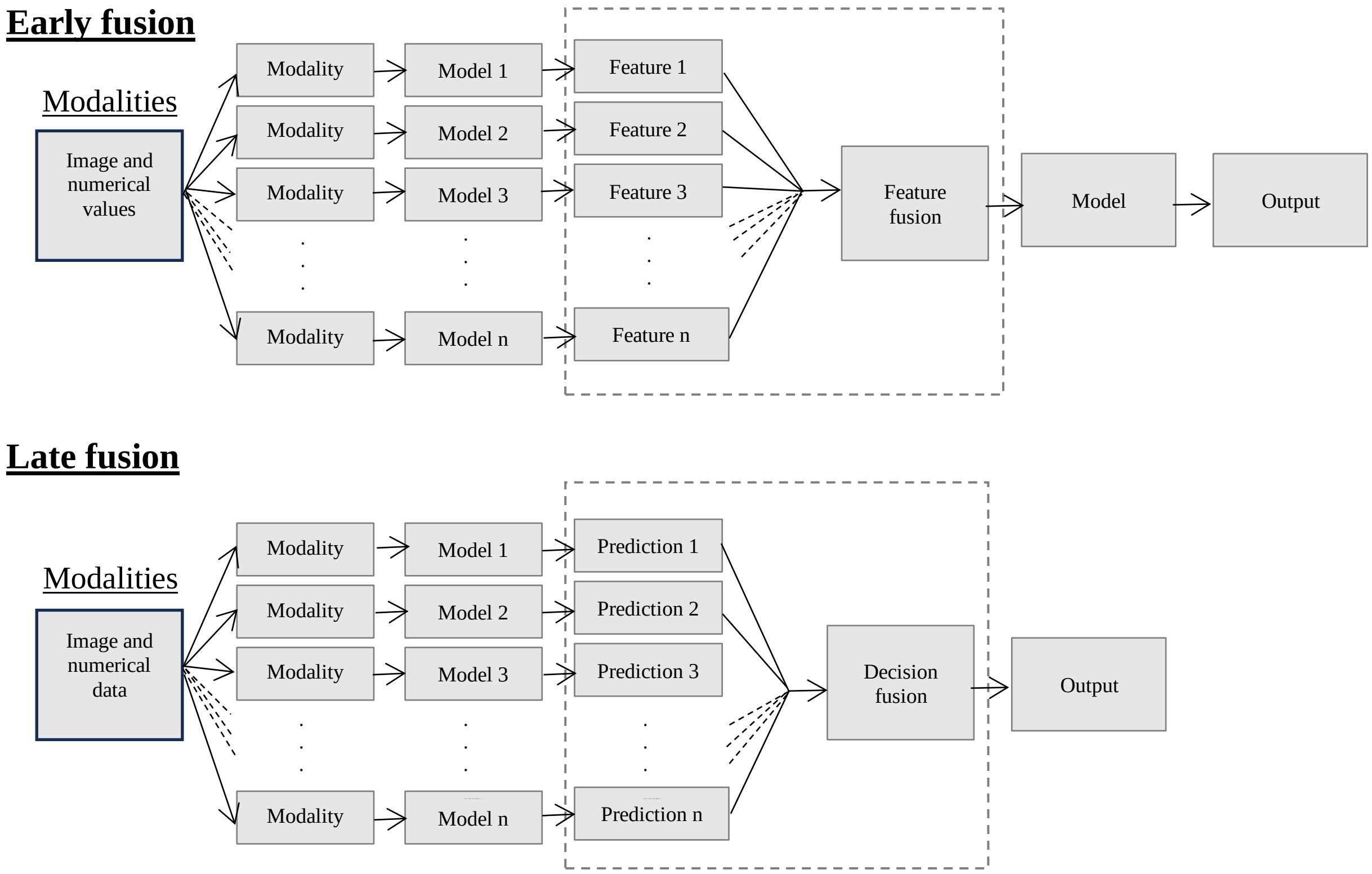

Fig 6. The general schematic of early and late fusion techniques in multi-modality methods.

Most studies [50,54] in this review have employed an early fusion approach, allowing DL models to learn from each modality independently before combining their outputs.

## 3.2. Processing strategies

Two main processing strategies are commonly used in DL-based glaucoma detection studies: single-step (end-to-end) and two-step approaches (Fig 7) [11]. The single-step strategy processes input from various data types to classify glaucoma directly. This approach integrates all necessary processes into one continuous operation, ignoring the need for intermediate steps.

Conversely, the two-step strategy employs a structured, hierarchical approach. It begins with the detailed segmentation of anatomical features, such as OD, OC, and retinal layers. Following the segmentation, the approach proceeds to the classification phase, where the segmented images are analyzed for glaucoma classification. As a result of the latter method, each phase of the detection process can be divided into discrete steps within the DL pipeline.

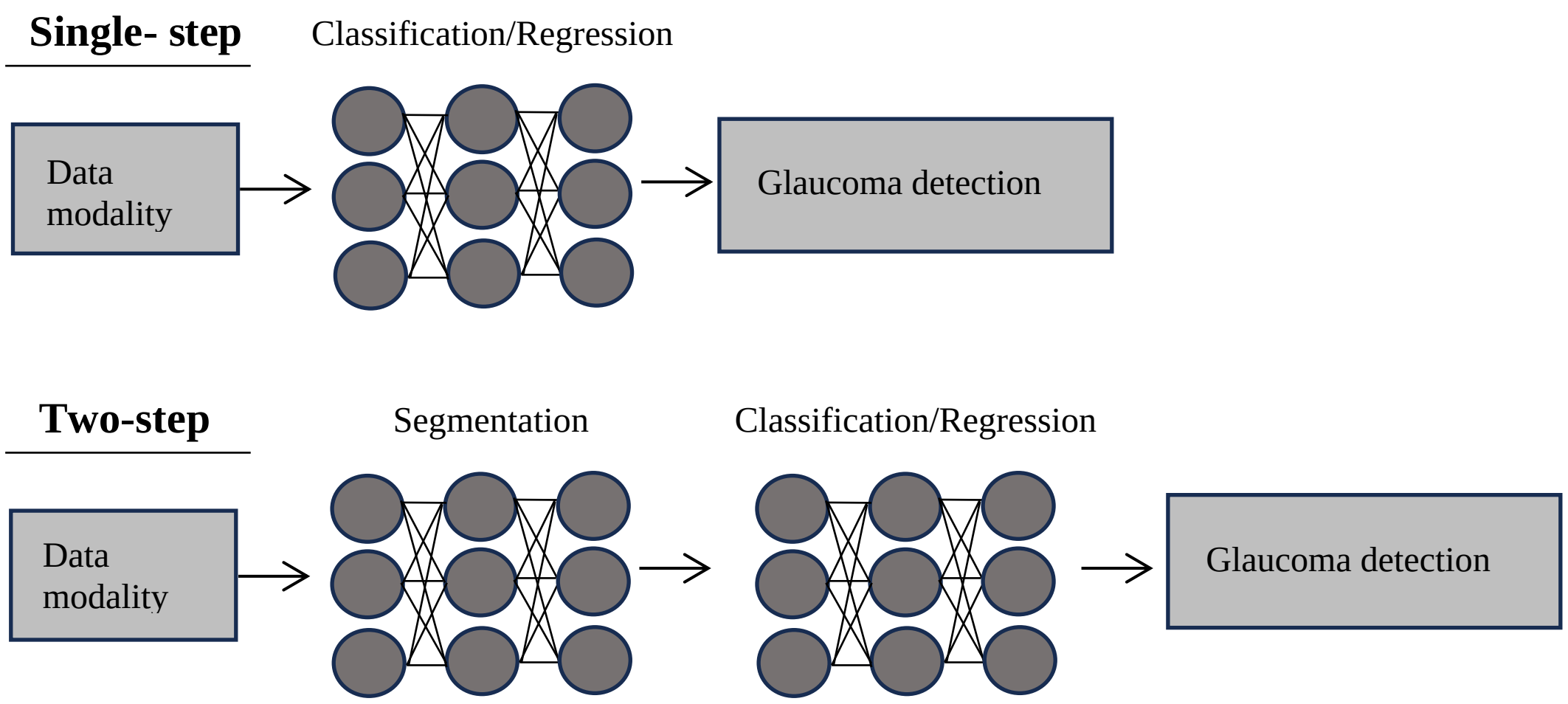

Fig 7. The processing strategies.

.

Reflecting on the evolution of these strategies, Table 3 demonstrates a clear trend towards the increasing adoption of single-step strategy [8,49,55–58]. The overall count of studies using this strategy rose from one in 2016 to 40 in 2023. Despite the main role played by the single-step strategy, two-step studies [59–63] are still present in the field.

Table 3. Annual trend of different processing strategies in DL-based glaucoma detection studies.

| | Year | | | | | | | | |
|---|---|---|---|---|---|---|---|---|---|
| **Strategy** | 2016 | 2017 | 2018 | 2019 | 2020 | 2021 | 2022 | 2023 | **Total** |
| Single-step | 1 | 2 | 5 | 8 | 13 | 19 | 19 | 40 | **107** |
| Two-step | - | 1 | 1 | 4 | 6 | 8 | 11 | 15 | **46** |
| Total | 1 | 3 | 6 | 12 | 19 | 27 | 30 | 55 | **154** |

## 3.3. Models

Traditional machine learning methods for glaucoma detection initially relied on the manual extraction of features from retinal images [64–66]. However, the introduction of DL models markedly enhanced the analysis of complex retinal structures for detecting glaucoma, transitioning from handcrafted to automatic feature extraction. The study by Asaoka et al. [67] marked a significant milestone by employing a deep feed-forward neural network using VF data to distinguish pre-perimetric open-angle glaucoma patients from healthy individuals. In subsequent studies, other DL models, such as CNNs, GANs, and attention-based models, were developed to improve glaucoma detection efficiency.

Table 4 illustrates the application of these models to glaucoma detection, including classification, progression prediction, and image synthesis. As indicated in Table 4, CNNs are the most widely used models for glaucoma prediction, having been used in 108 studies for classification and 20 studies for predicting progression. GANs were used to classify images in four studies and used to synthesize images in three studies. Attention-based models have been evaluated in a few studies, with 17 using them for glaucoma classification and two for glaucoma progression prediction.

Table 4 Comparative analysis of DL models in different applications.

| | **Application** | | |
|---|---|---|---|
| **Deep learning models** | Classification | Progression prediction | Image synthesis |
| CNNs | 108 | 20 | - |
| GANs | 4 | - | 3 |
| Attention(s) | 17 | 2 | - |

### 3.3.1. Convolutional Neural Networks (CNNs)

**Background and overview:** CNNs [68] have emerged as a leading approach in DL after their success in the ImageNet competition [69]. CNNs automatically and adaptively learn spatial hierarchies of features, from low- to high-level patterns [70]. A CNN architecture typically includes convolutional layers for feature extraction, pooling layers to reduce dimensionality, and fully connected layers for classification. Fig 8(a) illustrates a standard CNN architecture used in glaucoma classification.

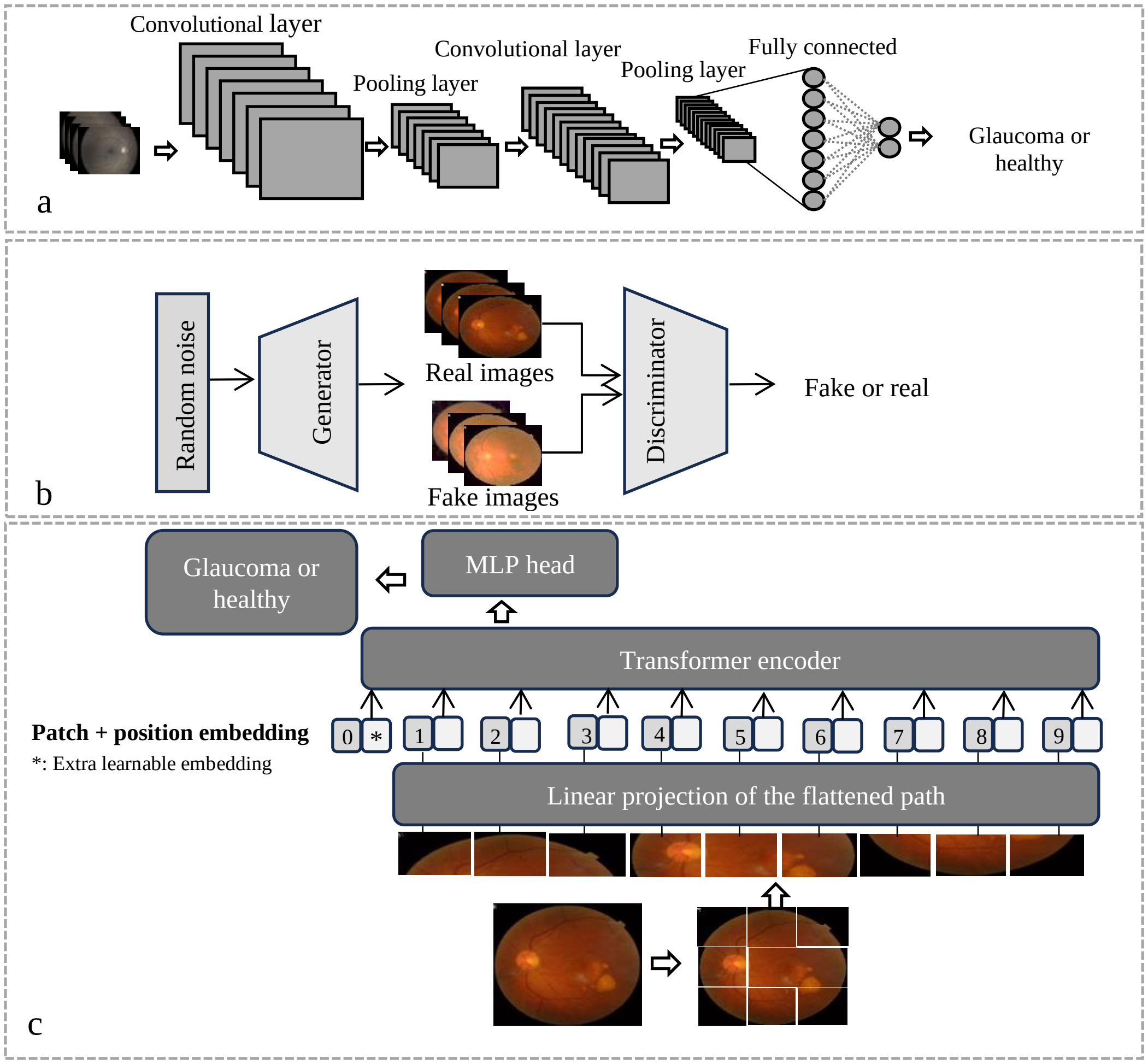

Fig 8. DL models architectures: (a) CNN, (b) GAN, (c) ViT.

**Classification**: CNNs have demonstrated remarkable performance in various single and two-step glaucoma classification tasks, including binary classification (e.g., glaucoma or normal) [6–10,54,71,72], multi-class classification of glaucoma stages (e.g., mild, moderate, severe) [9,73–75], multi-class classification of glaucoma among other retinal diseases (e.g., diabetic retinopathy, macular degeneration, cataracts, and glaucoma) [76–83] and classification of glaucoma subtypes (e.g., open-angle and angle-closure) [34,36,84].

Studies evaluating DL models for glaucoma classification are mostly based on single-step strategies. One notable example is MVGL-Net [85], a multi-view learning framework for advanced angle-closure diagnosis. MVGL-Net achieved an AUC of 0.71 and demonstrated potential for real-world applications through its high performance on external datasets. In addition, the application of geometric DL techniques [32,86], such as PointNet and dynamic graph CNN, in glaucoma classification has offered efficient representations of complex 3D structures like the ONH. Furthermore, multimodal models have also shown a large degree of promise for glaucoma classification [73,87–90]. For example, Mehta et al. [87] combined Densenet, InceptionResnetV4, and decision trees with OCT, color CFP photos, and clinical data, achieving an AUC of 0.97 for glaucoma classification. Demir et al. [83] developed an R-CNN and LSTM-based algorithm that achieved a high degree of accuracy, sensitivity, and specificity across eight ocular diseases, including glaucoma, using the ODIR dataset.

In some studies [91–101], classification-based glaucoma detection involves a two-step strategy. Researchers first segment anatomical structures, such as OC, OD, and blood vessels, and then use the predicted segmentations in the classification task. Examples of segmentation masks for these anatomical structures are shown in Appendix A (Fig A.2). The state-of-the-art segmentation methods frequently used for OD, OC, and vessel segmentation in the glaucoma context are U-Net [102], DeepLab [103], and Segnet [104]. These two-step glaucoma classification studies [105,106] often incorporate additional preprocessing and post-processing steps to enhance the segmentation accuracy. For instance, Fu et al. [105] used a modified U-Net (M-Net) with polar transformation as a post-processing step for improving OD/OC segmentation. Fu et al. achieved an AUC of 0.89, the highest of all studies evaluated here.

However, to mitigate the requirement for extensive pre- and post-processing steps and to address the high computational demands of DL models, in certain studies [107,108], alternative methods have been proposed. Tabassum et al. [108] introduced the Cup Disc Encoder-Decoder Network (CDED-Net), a model characterized by a shallower network structure designed to reuse information at the decoder stage. This approach aimed to streamline the model and improve efficiency without compromising performance. CDED-Net achieved an average AUC of 0.96 for glaucoma classification.

In addition, several studies [76,109–111] have investigated other anatomical structure segmentation, such as blood vessels [76,109] and retinal layer [111], followed by CNN-based classification. For instance, Panda et al. [30] developed a U-Net model to segment neural and connective tissues in the ONH using OCT images and achieved a glaucoma detection accuracy of 92.0%.

**Progression prediction:** Glaucoma, recognized as a progressive disease, necessitates continuous monitoring to manage it clinically. CNNs have emerged as pivotal tools in assessing glaucoma progression, offering detailed analyses through various indicators. These include VF sensitivity indices [31,46,51,112,113], VF maps [42], specific VF parameters (e.g., mean VF deviations, PSD) [114–116], as well as RNFL thickness [57,117–119]. As such, much work has been devoted to developing CNN-based approaches using these modalities. Hashimoto et al. [31] utilized a pattern-based regularization CNN to predict VF threshold sensitivity. They trained their network on the thickness of specific macular layers and achieved an average absolute error of 2.84 ± 2.98 dB across the entireMultimodal approachesoaches have also shown bene resultsficial; Dixit et al. [49] assessed glaucoma progression using a convolutional LSTM neural network with both VF and clinical data, achieving an AUC of 0.89. Similarly, Hussain et al. [120] developed a model for predicting glaucoma progression using a DL framework trained on OCT images, VF values, and demographic/clinical data, achieving an AUC of 0.83 for predicting glaucoma progression 6 months in advance.

### 3.3.1. Generative Adversarial Networks (GANs)

**Background and overview:** GANs are a deep neural network architecture type that can generate new samples from a given probability distribution (Fig 8(b)) [121]. A GAN is comprised of two components: a generator and a discriminator. Using a random noise vector as input, the generator creates fake data to mimic a real data distribution. The discriminator, functioning as a binary classifier, attempts to differentiate real data from the generator's fake data. The algorithm converges when the discriminator can no longer distinguish between real and generated data [122].

Additionally, GANs are used for domain adaptation to bridge the distribution gap between different datasets, enhancing model performance on new unknown data sources. We identified three studies [120,123] that used GANs to synthesize retinal images, while other studies have focused on employing GANs to tackle domain-related challenges in the segmentation of OC and OD for glaucoma classification [124–126].

**Classification**: Studies on OD and OC segmentation using CNN models often face challenges with new datasets due to domain shift issues. GAN-based studies in glaucoma detection employ unsupervised domain adaptation methods to address this issue [124–126]. For instance, Wang et al. [126] developed a patch-based adversarial learning framework (pOSAL) for segmenting OD and OC from various CFP images using images from open-source datasets such as Drishti-GS, RIM-ONE-r3, and REFUGE. Wang et al. [126] assessed the effectiveness of their approach using Dice coefficients and CDRs for glaucoma screening. Similarly, Liu et al. [125] introduced an unsupervised domain adaptation method, the Efficient Classification and Segmentation Network, based on Depth Domain Adaptation (ECSD). This approach was considered for simultaneous OD and OC segmentation and glaucoma classification. This method underwent testing on the Drishti-GS, RIM-ONEr3, and REFUGE datasets and showed its efficiency in reducing the impact of domain shift on segmentation tasks while improving the accuracy of glaucoma screening.

**Image synthesis:** Diaz-Pinto et al. [127] developed a deep convolutional GAN to create a retinal fundus image synthesizer and semi-supervised learning method for glaucoma assessment using Deep Convolutional Generative Adversarial Network (DCGAN). The method accurately distinguished between glaucomatous and normal images, achieving an AUC of 0.90. Furthermore, in some studies [120,123], GANs were used to synthesize OCT images from CFP data, facilitating early diagnosis without expensive OCT equipment. For example, Chang et al. [123] developed a glaucoma detection system using a GAN for generating OCT images from CFPs. Experimental results indicated a 97.8% similarity between generated and real OCT images, with the classification model achieving an 83.17% accuracy.

### 3.3.2. Attention (s)

**Background and overview**: The attention mechanism was first introduced in Natural Language Processing (NLP) tasks for machine translation tasks [128,129]. Attention has been adapted in computer vision and has become an important component of neural network architectures due to how it mimics the human mechanism for reasoning about visual stimuli by focusing on important parts of the input [130].

The Vision Transformer (ViT) model [131] is an attention-based model commonly used in computer vision tasks. In ViTs, illustrated in Fig 8(c), input images (e.g., CFP images) are split into fixed-size patches, transformed into embeddings, and processed through a transformer encoder using self-attention. This encoder enhances image understanding by contextualizing each patch's relation to others. The resulting aggregated embedding is fed into a Multi-Layer Perceptron (MLP) head for downstream classification [132].

These attention-based methods have been applied in various aspects of glaucoma detection, including glaucoma classification and progression prediction. In the following sections, we explore how attention-based methods have contributed to each of these applications.

**Classification:** Studies on attention-based classification in glaucoma have primarily utilized single-step classification tasks [133–142]. For instance, Li et al., [143], developed an Advanced Glaucoma Convolutional Neural Network (AG-CNN) model, incorporating three key components: an attention prediction subnet, a pathological area localization subnet, and a glaucoma classification subnet. This approach allowed the model to focus on critical areas within the images, enhance the localization of pathological features, and accurately classify the presence of glaucoma. The AG-CNN model introduced by Li et al. [143] achieved an AUC of 0.96. Similarly, Xu et al. [134] introduced the Transfer Induced Attention Network (TIA-Net) for glaucoma detection. This approach involved transferring knowledge from a related ophthalmic dataset to enhance the model's detection capabilities. Benefitting from transfer learning, TIA-Net effectively learned relevant features for glaucoma detection and achieved an AUC of 0.93.

Furthermore, Garcia Pardo et al. [136] developed a model combining a residual architecture and attention block for feature extraction, followed by a CNN and LSTM-based predictive model. Using SD-OCT volumes, they achieved an AUC of 0.88. Song et al. [50] developed the Deep Relation Transformer (DRT) model for glaucoma detection. DRT was a transformer-based multimodal approach combining VF and OCT data. It employed a global relationship module, a guided regional relationship module, and an interaction transformer module for effective information extraction. The model achieved an accuracy of 88%.

The attention-based method has become increasingly popular in glaucoma detection when using a two-step strategy. These methods effectively address challenges like blurred boundaries between the OC and OD and overlapping blood vessels with the OC [144]. For example, Bhattacharya et al. [145] introduced PY-Net, an OD/OC segmentation method using an attention module. It combined receptive field blocks, attention modules, a densely connected spatial pyramid decoder, and multi-scale coarse segmentation maps to achieve a mean average error of 0.02 for disease identification. Furthermore, Zhou et al. [146] proposed EARDS, an EfficientNet-based model for OD and OC segmentation and vCDR calculation. EARDS used an EfficientNet-b0 encoder, attention gate, residual depth-wise separable convolution blocks, and a novel decoder network to achieve AUC scores of 0.97.

**Progression prediction:** Monitoring glaucoma progression is challenging due to performance variability and the absence of a standard method [147]**.** Glaucoma's progression is marked by subtle changes in the optic nerve and RNFL. Thus, attention-based methods are promising in this context, offering precise analysis of these subtle changes. A few studies, including Hou et al. [148], used attention-based methods to predict glaucoma progression. Their Gated Transformer Network (GTN) analyzed OCT images to detect visual field deterioration, combining trend-based and event-based methods. With at least five OCT scans per case, the GTN achieved a 0.97 AUC. Furthermore, Hu et al. [149] developed GLIM-Net, a transformer model predicting glaucoma likelihood using irregularly sampled CFP images with an AUC of 0.93. Uniquely, they extended their model to predict specific future times, achieving an accuracy of 89.5%.

## 4. Discussion

To our knowledge, there has been no comprehensive review of DL-based glaucoma detection studies that simultaneously examine model architectures, applications, processing approaches, input data modalities, and the evolution of these aspects since the introduction of DL in this field. Expanding upon a comprehensive analysis of the progress in DL for glaucoma detection, our discussion will delve into these critical facets in greater detail. Moreover, we aim to highlight the prevailing challenges within these areas and propose potential avenues for future exploration and research.

### 4.1. Discovered patterns and trends

CFP imaging has gained significant popularity in DL-based glaucoma detection research, primarily because of its compatibility with DL algorithms [150] and the widespread availability of publicly accessible data that enable comprehensive training and testing of neural networks [92,126,151,152]. In addition, recent studies have shown an increased use of OCT in glaucoma detection, driven by OCT's ability to deliver high-resolution cross-sectional images of the retina. The commercial availability and clinical acceptance of OCT for diagnosing retinal diseases [153] further contribute to OCT's incorporation into research. The trend towards using VF testing in glaucoma studies aligns with its clinical importance for assessing glaucoma's functional impact. Research emphasis on VF data has evolved, initially favoring VF, then shifting towards imaging technologies like CFP and OCT for their detailed structural insights. This reflects a growing emphasis on integrating functional and structural data to enhance diagnostic precision, underscored by the increased trend in multimodal studies [52].

Some studies have used private datasets for glaucoma detection. The utilization of the private dataset might be because 1) publicly available datasets fmultimodal imagingultimodalimaging are almost inaccessible, and 2) private datasets provide detailed analysis of glaucoma severity stages [154,155] and demographic [87,120] variations that public data may underrepresent. Utilizing both public and private datasets in model development may help improve generalizability to real clinical settings [156,157].

Even though there is an increasing trend in using single-step and two-step strategies, there are several advantages and disadvantages to both these strategies in glaucoma detection. The single-step strategy is gaining popularity, likely due to advancements in DL that enable accurate classification from complex inputs without intermediary segmentation steps. These methods are less computationally expensive and well-suited for real-time clinical applications. Potentially, at the cost of increased resource requirements, the two-step strategy can offer a detailed analysis of ocular structures, allowing clinicians to identify minute changes indicating glaucoma progression. The two-step strategy benefits from incorporating domain-specific knowledge, like ONH anatomy, potentially enhancing detection accuracy and providing insights into the model's decision-making process. However, two-step strategies are complex and resource-intensive, and their sequential design can be challenging to implement and train. Furthermore, their effectiveness heavily relies on accurate initial segmentation. Inaccurate segmentation can lead to erroneous feature extraction, compromising the analysis' reliability and increasing the risk of misclassification.

## 4.2. Gaps and Future Directions

The lack of publicly available datasets, especially for OCT, VF, and multimodal data, poses a significant difficulty in the development of DL-based glaucoma detection models. Although some recent datasets for OCT [158–160] and multi-modality [161–163] have been released, there remains a demand for additional contributions to gather and disseminate such data. Furthermore, given the substantial costs and expertise involved in data collection, investigating alternative training methods such as zero or few-shot learning [164] could be beneficial in addressing the challenges posed by limited dataset availability.

Another challenge is the lack of standardized data labeling criteria. Glaucoma is a complex disease, and various structural and clinical evaluations are needed, including CDR, intraocular pressure levels, VF defects, and RNFL thickness. The reviewed studies used diverse parameters for glaucoma labeling, highlighting the need for standardized guidelines to ensure consistency and reduce ambiguity in glaucoma research. Ideally advanced by field experts, this standardization is critical for enhancing glaucoma detection methods and improving patient outcomes.

There are also challenges associated with datasets due to their intrinsic bias, which is a result of different protocols for data collection across institutions, differences in recording devices, and disparities among subpopulations with diverse demographic profiles. This intrinsic bias has the potential to undermine the generalizability of glaucoma detection with DL models due to domain shift issues. Thus, it highlights the importance of collaboration among institutions to train DL models using a variety of datasets. As a result, such collaborations should take into account concerns regarding patient privacy and data security. These concerns can be addressed through techniques such as de-identification and advanced algorithms such as Federated Learning (FL) [165]. It should be noted that only one study in this survey focused on data privacy [166]. Given the importance of this area, more research efforts should be devoted to this field in the future.

In addition to collaborative efforts aiming to improve the performance of DL-based glaucoma detection models, it is crucial to ensure their reliability, particularly when applied in new, unknown environments. The reliability of these model,s sometimes called the confidence level, can be assessed through uncertainty estimation techniques [167], such as Bayesian neural networks [168], confidence calibration [169], and ensemble methods [170]. Despite the importance of these evaluations, only one study we reviewed has specifically assessed their reliability for glaucoma detection [171]. This gap highlights the need for further investigation into the robustness of DL models in the glaucoma detection field.

Moreover, standardized metrics are needed to evaluate the glaucoma detection model in the future. The studies reviewed here have used a range of metrics to evaluate DL-based glaucoma detection algorithms, making direct comparisons of these studies difficult. To compare the performance of studies surveyed in this review, we looked at those that reported AUROC. As detailed in Appendix B (Table B.2), mean Area Under the Receiver Operating Characteristics (AUROC) values for DL-based glaucoma detection studies have steadily risen from 2016 to 2020, remaining relatively steady through 2023. Additionally, Appendix B (Table B.3) emphasizes that some studies, such as Lee et al. [172] and Vijaya Kumar et al. [74], have achieved AUROCs as high as 1.00. However, comparing these models remains challenging without a standardized evaluation system, underscoring the need for benchmarks in DL-based glaucoma detection.

## 5. Conclusion

In conclusion, this review paper highlights significant developments in glaucoma detection using DL, guided by the PRISMA framework. We have investigated advancements in data modalities, processing techniques, model architectures, and applications. Despite these developments, there are still challenges, such as data diversity and the reliability of the models. Future research should focus on overcoming these hurdles.

## Declaration of competing interest

The authors declare that they have no known competing financial interests or personal relationships that could have appeared to influence the work reported in this paper.

# Appendix. A

**Fig A.1: Histogram of the citation frequency by year for DL-based glaucoma detection studies reviewed in this survey**.

This figure illustrates the distribution of citation counts for deep learning (DL)-based glaucoma detection studies, grouped by their publication year from 2018 to 2022. Due to the emerging stage of research in those initial years, we excluded the years 2016 and 2017 from this analysis, as well as 2023, due to its recent publication status.

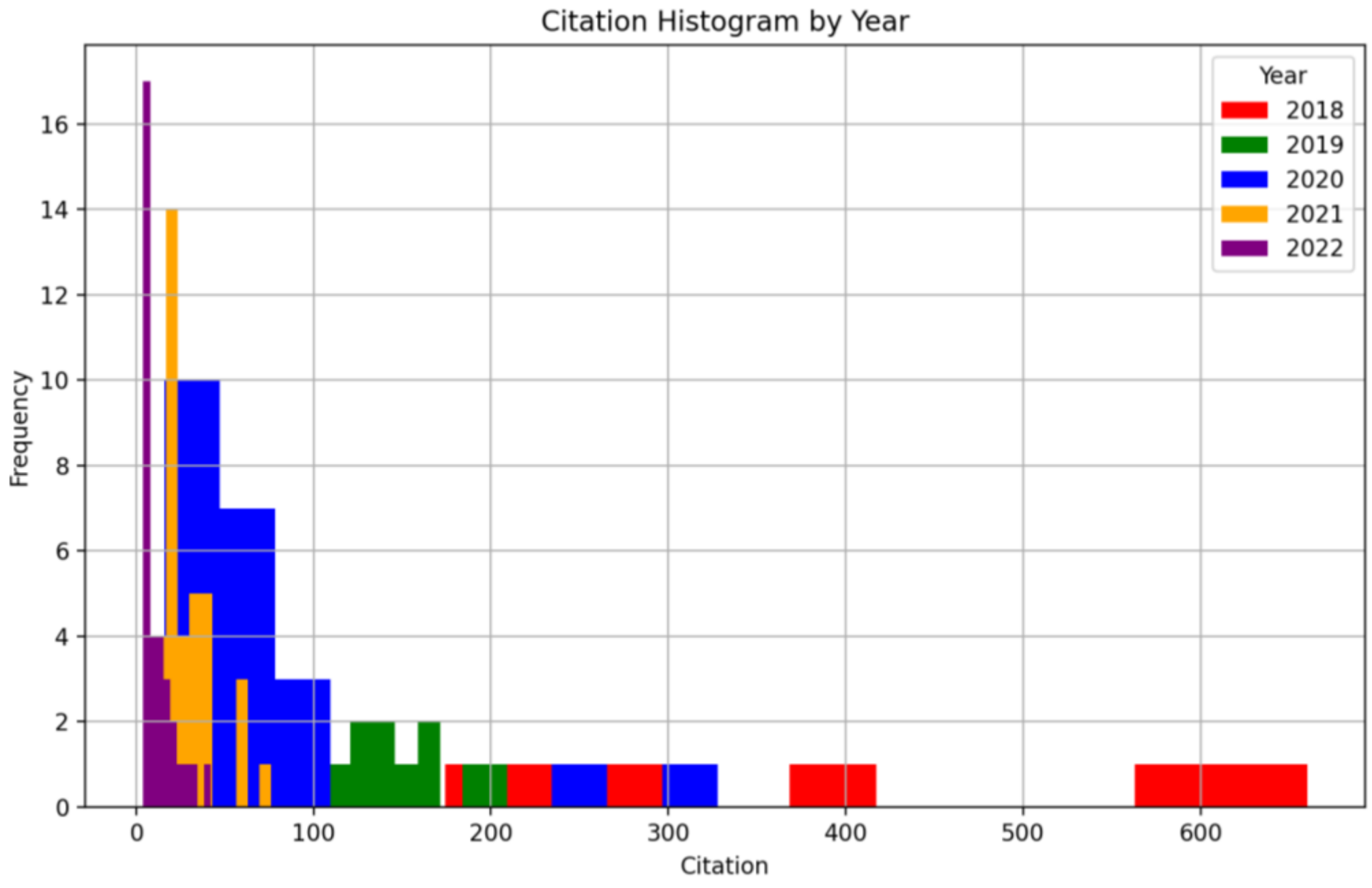

**Fig A.2: Segmentation masks for OC, OD, and blood vessels**.

Anatomical structure segmentation (a) blood vessel, and (b) OC and OD. The segmentation of blood vessels is sourced from the publicly accessible FIVES dataset [173], while the segmentation of the OC and OD is derived from the Composite Retinal CFP and OCT Dataset [174,175].

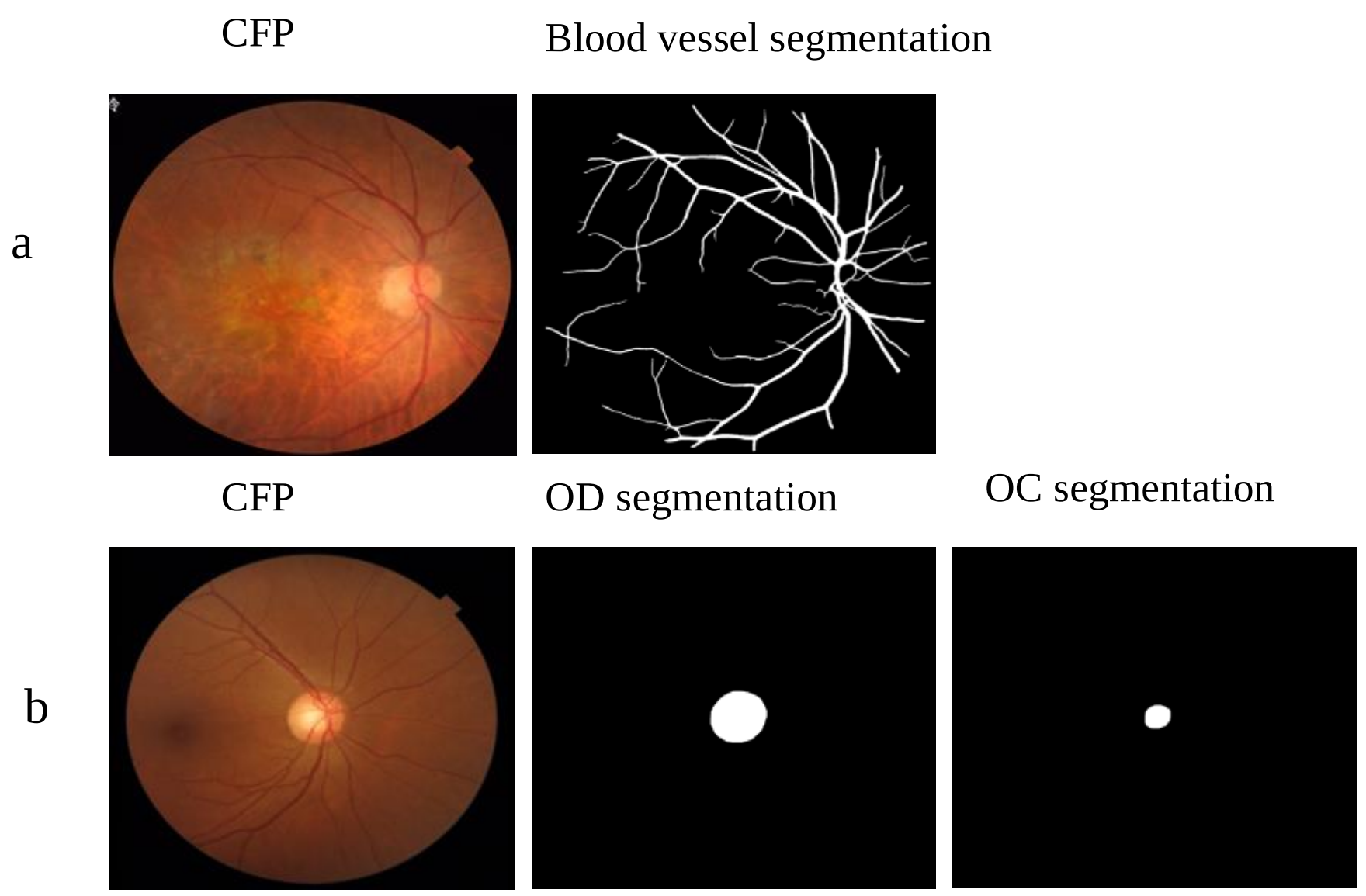

# Appendix. B

**Table B. 1** Glaucoma detection dataset: publicly accessible or available upon registration.

| Dataset | Modality | Number of data | Application | Accessibility | Link |
|---|---|---|---|---|---|
| **ACRIMA** | Fundus | Total: 705<br>Glaucoma: 396<br>Normal: 309 | Classification | Public | link |
| **KEH (Kim's Eye Hospital)** | Fundus | Total: 1544<br>Advanced_glaucoma:467<br>Early_glaucoma: 289<br>Normal: 788 | Classification | Public | link |
| **HRF** | Fundus | Total: 45<br>Glaucoma: 15<br>Normal: 15:<br>DR: 15 | Classification<br>Segmentation of the field of view (FOV)<br>Segmentation of the blood vessels | Public | link |
| **FIVES** | Fundus | Total: 800<br>DR: 200<br>Glaucoma: 200<br>AMD: 200<br>Normal: 200 | Classification<br>Segmentation of the blood vessels<br>Image quality assessment | Public | link |
| **G1020** | Fundus | Total: 1020<br>Healthy: 724 | Classification | Public | link |
| | | Glaucoma: 296 | Segmentation of the OC and OD | | |
| **DRISHTI-GS1** | Fundus | Total: 101<br>Normal: 31<br>Glaucoma: 70 | Classification<br><br>Segmentation of the OC and OD | Public | link |
| **EyePACS AIROGS** | Fundus | Train set:<br>Total: 101,442<br>Referable glaucoma: 3,270<br>No referable glaucoma: 98,172<br><br>Test set:<br>Total: 11,290<br>Referable glaucoma: 1,602<br>No referable glaucoma: 8,134<br>Ungradable: 1,554 | Classification | Public | link |
| **BEH** | Fundus | Total: 634<br>Normal: 463<br>Glaucoma: 171 | Classification | Public | link |
| **DR-HAGIS** | Fundus | Total: 40<br>Glaucoma: 10<br>Hypertension:10<br>DR: 10<br>AMD: 10 | Classification<br><br>Segmentation of the blood vessels | Public | link |
| **Chaksu-IMAGE** | Fundus | Total: 1345 | Segmentation of the OC and OD | Public | link |
| **AGE** | AS-OCT | Total: 4800<br>Angle-closure: 960<br>Open-angle: 3840 | Scleral spur localization<br><br>Classification | Public upon registration | link |
| **BIOMISA** | Funds and OCT | Total: 50 (normal, glaucoma, or suspect based on four different experts' opinion) | Classification | Public | link |
| **GAMMA** | Fundus and OCT | Total: 300<br>Non-glaucoma: 100<br>Early-glaucoma:100<br>Progressive-glaucoma: 100 | Classification<br><br>Segmentation of the OC and OD<br><br>Localization of the fovea | Public upon registration | link |
| **CRFO** | Funds and OCT | 9,268 OCT and 180 fundus scans | Classification | Public | link |
| **GRAPE** | VFs, fundus, OCT measurement, and clinical information | Total: 263<br>Open-angle glaucoma (OAG): 254<br>Angle-closure glaucoma (ACG): 9 | Classification<br><br>Segmentation of the OD | Public | link |
| **Harvard-GD500** | RNFLT maps and visual field mean deviation (MD) | Total: 500 | Classification | Public upon registration | link |

**Table B. 2** Statistical analysis of the AUROC of the glaucoma detection studies surveyed in this study.

| Year | 2016 | 2017 | 2018 | 2019 | 2020 | 2021 | 2022 | 2023 |
|---|---|---|---|---|---|---|---|---|
| AUROC | | | | | | | | |
| Range (min,max) | 0.92 | (0.94,0.94) | (0.89, 0.98) | (0.85, 0.99) | (0.80, 1.00) | (0.80,0.99) | (0.84,0.99) | (0.86,1.00) |
| Mean+STD | 0.92 | 0.94±0.002 | 0.94±0.036 | 0.94±0.042 | 0.95±0.050 | 0.93±0.057 | 0.94±0.052 | 0.95±0.036 |

**Table B. 3** Summary of top ten studies reported best AUROC.

| Reference | Modality | Data | Dataset name or total number of data (if data is private) | Model | Performance<br>AUC | Description |
|---|---|---|---|---|---|---|
| **Lee et al., 2020** [172] | SD-OCT | Private | 282 | NASNet | Total: 0.99<br>Early glaucoma: 0.98<br>Moderate-to-Severe glaucoma: 1.000 | For each subject, 4 images were used as the input for the NASNet architecture, including (1) a GCIPL thickness map; (2) a GCIPL deviation map; (3) an RNFL thickness map; (4) an RNFL deviation map. |
| **Vijaya Kumar and Sharma, 2023** [74] | CFP | Public | ACRIMA<br>RIM-ONE<br>Harvard Dataverse (HVD)<br>Drishti | ResNet50<br>AlexNet<br>VGG19<br>DenseNet-201<br>Inception-ResNet-v2 | ACRIMA: 1.00<br>RIM-ONE: 0.95<br>HVD: 0.93<br>Drishti: 0.91 | The output of different CNN model was fused to make the final decision |
| **Z. Li et al., 2021** [176] | Ultra-widefield CFP (UWF) | Private | 22,972 | InceptionResNetV2 | 0.983–0.999 | - |
| **Prabhakaran et al., 2022b** [177] | SD-OCT | Private | 2,154 | Depthwise Separable Convolution (DSC) | 0.99 | - |
| **Liu et al., 2019** [9] | CFP | Private | 274,413 | ResNet | 0.99 | - |
| **Ran et al., 2023** [166] | 3D OCT | Private | 9,326 | 3D DenseNet121<br>3D ResNet10<br>3D ResNet18 | 3D-DenseNet121: 0.794–0.991<br>3D-ResNet10: 0.809–0.991<br>3D-ResNet18: 0.794–0.992 | Federated learning (FL) paradigm was used in this study. |

**Table B. 3** Summary of top ten studies reported best AUROC (Continued).

| Reference | Modality | Data | Dataset name or total number of data (if data is private) | Model | Performance | Description |
|---|---|---|---|---|---|---|
| **Hemelings et al., 2019** [178] | CFP | Private | 8,433 | ResNet-50 | 0.99 | Active learning process was used in this study. |
| **Shamsan et al., 2023** [179] | CFP | Public | Oculur Recognition<br><br>IDRiD<br><br>HRF | DenseNet121<br><br>MobileNet | MobileNet: 0.93<br><br>DenseNet121: 0.94<br><br>Fused MobileNet and DenseNet121: 0.97<br><br>Fused MobileNet and handcrafted: 0.99<br><br>Fused DenseNet121and handcrafted: 0.99 | The fusion of deep features of DL models were fused by handcrafted features. |
| **Feng Li et al., 2020** [73] | CFP+ medical history data | Private | 26,585 | ResNet101 | 0.992 | - |
| **Akter et al., 2022** [110] | OCT | Private | 200 | ResNet18<br><br>VGG16,<br><br>A custom CNN with 24 layers | ResNet18: 0.99<br><br>VGG16: 0.99<br><br>A custom CNN with 24 layers: 0.99 | - |